\documentclass[twocolumn]{aa}

  \usepackage{amscd}
  \usepackage{amsmath}
  \usepackage{amssymb}
  \usepackage[utf8]{inputenc}
  \usepackage[T1]{fontenc}  
  \usepackage[varg]{txfonts}
  \usepackage{graphicx}
  \usepackage{natbib}
  \usepackage{verbatim}
  \usepackage{array}
  \usepackage{isotope}
  \usepackage{units}
  \usepackage{subfigure}

\begin{document}

\title{C+O detonations in thermonuclear supernovae: Interaction with
  previously burned material}
\titlerunning{C+O detonations in thermonuclear supernovae}

\author{A.\ Maier \and J.C.\ Niemeyer}
\institute{Lehrstuhl für Astronomie, Universität Würzburg, Am Hubland, D-97074
  Würzburg, Germany}

\abstract{
In the context of explosion models for Type Ia Supernovae, we present
one- and two-dimensional simulations of fully resolved detonation 
fronts in degenerate C+O White Dwarf matter including clumps of
previously burned material. The ability of detonations to survive the
passage through sheets of nuclear ashes is tested as a function of the
width and composition of the ash region. We show that detonation fronts 
are quenched by microscopically thin obstacles with
little sensitivity to the exact ash composition. Front-tracking
models for detonations in macroscopic explosion simulations need to
include this effect in order to predict the amount of unburned
material in delayed detonation scenarios. 
  
}

\maketitle

\section{Introduction}
Driven by the ever increasing amount and quality of observational data
for Type Ia Supernovae (SNe Ia) \citep[e.g.,][]{Lea05,Mea05} and their continuing role
as cosmological distance indicators \citep{Rea05}, the theoretical efforts to
understand the physics of exploding C+O White Dwarf stars near the
Chandrasekhar mass have progressed significantly. We now have at our
disposal three-dimensional hydrodynamical models including different
algorithms to track thermonuclear flames, detonations, and unresolved
turbulence \citep[e.g.,][]{GKO04b,RGRe05}, as well as the 
results of fully resolved Rayleigh-Taylor unstable flame fronts
showing the transition to isotropic turbulence on small scales
\citep{ZWRe05}. Nevertheless, no consensus has been reached on whether pure
turbulent deflagrations can explain SN Ia observations as indicated by
high-resolution multi-point ignition models or
whether delayed detonations of some sort are needed to burn more
material at low densities. 

It is undisputed, however, that any detonation must be preceeded by a
phase of subsonic burning that pre-expands the star and, as a result of
the Rayleigh-Taylor (RT) instability, leaves behind highly inhomogeneous
regions of burned and unburned material. Large, merging RT mushrooms
frequently produce large volumes of fuel fully enclosed by ash (but not vice
versa). One of the key virtues of delayed detonations is their suspected
ability to deplete much of the remaining unburned C+O in the White
Dwarf's core region. Large enough quantities of such
material would produce low-velocity carbon lines in the nebular
spectra of SNe Ia whose observational absence can already rule out
coarsely resolved deflagration models with single-point central ignition
\citep{Kea05}. 

The work reported in this paper sidesteps the key issue for evaluating the
``naturalness'' of SN Ia scenarios that invoke a detonation: under
which microscopic conditions can a thermonuclear detonation in
degenerate C+O material be ignited \citep[e.g.,][]{NW97,KOW97,N99}? We instead
assume that a detonation 
has formed by some unspecified mechanism and begins to propagate
both outward and inward from its point of birth. Upon impacting on a
region of ash produced by the preceeding deflagration, thermonuclear
reactions cease to provide the overpressure that compensates for the
dissipation of the shock front. After propagating through the ash and
entering fresh fuel on the other side, the shock may either be
sufficiently powerful to re-ignite a detonation or not, in which case
fully enclosed pockets of fuel remain unburned. 

One aim of
this work was to numerically determine the linear size of a sheet of ashes
that a C+O detonation can still cross and survive\footnote{After completion
  of this work we were informed that a similar study had previously
  been performed by \cite{L04}.}. This was done with 1D, fully resolved simulations 
using the FLASH code \citep{fry00} that has previously been employed successfully to study
thermonuclear C+O detonations by \citet{Tea00}. Only for the lowest fuel
density in our parameter space, $\rho = 3.2 \cdot 10^7$ g cm$^{-3}$,
did we find successful detonation re-ignitions up to an ash thickness of
roughly 6 detonation widths (corresponding to 8 cm). In all other
cases, the detonation failed to re-ignite on the other side,
regardless of the detailed composition of the burned matter (pure
$^{56}$Ni, $^{28}$Si, or NSE).  

In addition, we carried out a small number of 2D simulations in order
to test the ability of detonations to pass through thin funnels of
fuel separating neighoring regions of ash. Not surprisingly, we found
that a minimum width of these funnels exists that allows the
detonation to pass if this width is close to the detonation
thickness. 

The structure of this paper is as follows. We describe the
numerical setup and calibration runs in Sect.\ 2, followed by a summary
of the 1D (Sect.\ 3.1) and 2D (Sect.\ 3.2) parameter
studies. Conclusions are presented in Sect.\ 4. , in particular those 
concerning the relevance of our results for large-scale simulations 
of detonations in exploding white dwarfs. 
 
\section{Tests of the numerical setup}
\label{sec:test}
The FLASH code is a modular, adaptive mesh, parallel simulation
code capable of handling general compressible flow 
problems in astrophysical enviroments. 
It uses an explicit, directionally split version of the
piecewise-parabolic method (PPM) \citep{col84} to solve the compressible
Euler equations and allows for general equations of state
using the method of \citet{col85}.
An equation of state appropriate for handling the 
interior of a C+O White Dwarf is implemented using a 
thermodynamically consistent table lookup scheme \citep{TS00} .
Source terms for thermonuclear reactions are solved using a 
semi-implicit time integrator coupled to a sparse matrix solver
\citep{Tim99}.
Further details about the algorithms
used in the code, the code's structure, and results of verification 
tests are documented in \citet{fry00} . 

As a good compromise between 
speed and precision, a 13 isotope $\alpha$-chain plus heavy-ion reaction 
network was used in our calculations. A network of this kind (called 
\verb|approx13|) using the isotopes \isotope[4]{He}, \isotope[12]{C},
\isotope[16]{O}, \isotope[20]{Ne}, \isotope[24]{Mg}, \isotope[28]{Si}, 
\isotope[32]{S}, \isotope[36]{Ar}, \isotope[40]{Ca}, \isotope[44]{Ti},
\isotope[48]{Cr}, \isotope[52]{Fe} and \isotope[56]{Ni} 
is implemented in FLASH and thorougly tested by
\citet{Tim99} and \citet{THW00}.

In order to quantify the necessary resolution for our simulation of 
detonation fronts with FLASH, we carried out a set of 1D test
simulations. The boundary conditions consisted of 
a system of length $l=\unit[1024]{cm}$ with a reflecting
boundary at $x=\unit[0]{cm}$ and an outflow boundary condition at
$x=\unit[1024]{cm}$. As initial conditions we chose a composition
of 50\% \isotope[12]{C} and 50\% \isotope[16]{O} at a temperature of 
$\unit[10^7]{K}$ and a material velocity of 
$v_x=\unit[0]{cm\ s^{-1}}$. To ignite the detonation
we set the inital conditions in an ignition area between $x=\unit[0]{cm}$ and
$x=\unit[25.6]{cm}$ to a temperature of 
$\unit[10^{10}]{K}$ and a material velocity of 
$v_x=\unit[10^9]{cm\ s^{-1}}$ to ensure sufficient numerical
diffusion for a successful detonation ignition. We examined
the speed and the nuclear energy generation rate of the
detonation for densities $\rho$ between 
$\unit[1.0 \cdot 10^{7}$ and $1.0 \cdot 10^{9}]{g\ cm^{-3}}$,
depending on the resolution of the simulation.
The speed of the detonation was evaluated by taking the derivative of the position 
for the maximum of the nuclear energy generation rate throughout the system
over time. We were able to reproduce the
results of \citet{Sha99} except for a difference of about 1\% that can
be accounted for by small differences in the employed reaction networks. 

As the detonation speed
is not very sensitive to the numerical resolution, we also
examined the magnitude of the maximum of the nuclear energy generation rate
for effective spatial resolutions of $\unit[0.4]{cm}$, $\unit[0.1]{cm}$,
$\unit[2.5 \cdot 10^{-2}]{cm}$, and $\unit[6.25 \cdot 10^{-3}]{cm}$.
\begin{figure}[t]
\centerline{
\includegraphics[width = \linewidth]
  {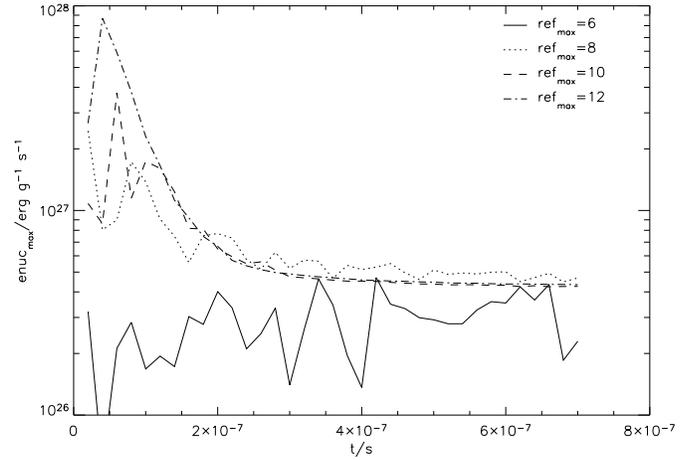}
}
\caption{Magnitude of the maximum of the nuclear energy generation rate
for a density $\rho = \unit[5.0 \cdot 10^{7}]{g\ cm^{-3}}$ and different spatial
resolutions $\unit[0.4]{cm}\ (\ \text{ref}_{max}=6) $ , 
$\unit[0.1]{cm}\ (\ \text{ref}_{max}=8)$,
$\unit[2.5 \cdot 10^{-2}]{cm}\ (\ \text{ref}_{max}=10)$, and 
$\unit[6.25 \cdot 10^{-3}]{cm}\ (\ \text{ref}_{max}=12)$}
\label{fig:nucmax}
\end{figure}
As can be seen in Fig.\ \ref{fig:nucmax}, the chosen inital conditions
generate a detonation with a high nuclear energy generation rate
that quickly converges to a stationary state. One can see that
increasing the resolution decreases the fluctuation amplitude of the maximum
of the nuclear energy generation rate up to the point
where the necessary resolution for a fully 
resolved detonation is reached. From this analysis
we found resolutions of $\unit[0.1]{cm}$, $\unit[2.5 \cdot 10^{-2}]{cm}$,
and $\unit[6.25 \cdot 10^{-3}]{cm}$ sufficient for fuel densities of
$\unit[3.2 \cdot 10^{7}]{g\ cm^{-3}}$, $\unit[3.2 \cdot 10^{7}]{g\
cm^{-3}}$, and $\unit[1.0 \cdot 10^{8}]{g\ cm^{-3}}$, respectively.
All further simulations were performed with these resolutions.

For a density of $\unit[1.0 \cdot 10^{7}]{g\ cm^{-3}}$, we found
that our initial conditions did not generate a stable detonation, and
for higher densities
$\rho = \unit[3.2 \cdot 10^{8} \dots 1.0 \cdot 10^{9}]{g\ cm^{-3}}$,
we were unable to reach the necessary resolution for a fully resolved
detonation.

\begin{figure}[t]
\centerline{
\includegraphics[width = \linewidth]
  {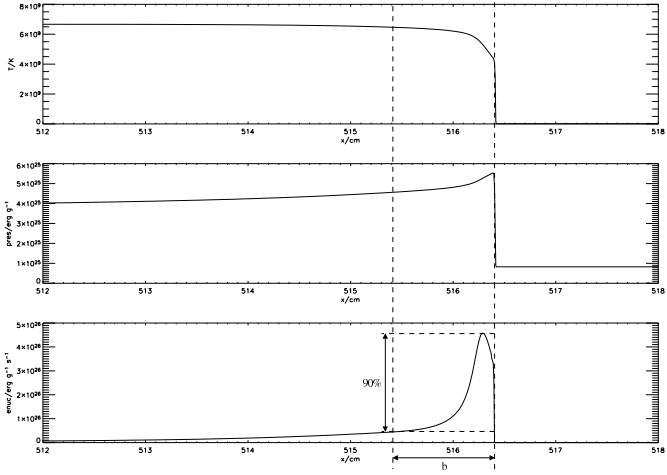}
}
\caption{Temperature, pressure, and nuclear energy generation profile 
of a detonation front for $\rho = \unit[5.0 \cdot 10^{7}]{g\ cm^{-3}}$
\ $\unit[4.0 \cdot 10^{7}]{s}$ after ignition. The dotted lines depict
the detonation width $b$ according to our definition.}
\label{fig:width}
\end{figure}

A convenient length scale to normalize our results is
given by the detonation width $b$, defined here as 
the distance
between the point of maximum pressure and the point where
the nuclear energy generation rate has dropped to 10\% of its 
maximum (Fig.\ \ref{fig:width}).
We found widths $b$ of $\unit[1.37]{cm}$, $\unit[1.00]{cm}$, and
$\unit[0.52]{cm}$  for fuel densities of
$\unit[3.2 \cdot 10^{7}]{g\ cm^{-3}}$, $\unit[5.0 \cdot 10^{7}]{g\
  cm^{-3}}$, and $\unit[1.0 \cdot 10^{8}]{g\ cm^{-3}}$.
Width-normalized lengths will be denoted henceforth by $l_s = l/b(\rho)$.

\section{Interaction of detonation fronts with ash obstacles}
\subsection{Properties of the ash obstacles}
The composition and thermal state of the nuclear ash 
produced by the deflagration front preceding the detonation depends
on the fuel density at the time it was consumed. In particular,
the ash composition can be expected to affect the behavior of the
detonation front since the shock energy is dissipated by the
dissociation of heavy nuclei. 

In order to study the sensitivity of shock dissipation to ash
composition, we examined five different setups (N, S, M1-M3) covering
a wide range of possible compositions (see Table\ \ref{tab:comp}).
\begin{table}[t]
\begin{center}
\begin{tabular}{|l|*{5}{c|}}
\hline
Isotop  & N             & S             & M1                                    & M2                                    & M3                                    \\
\hline
He4             &               &               & $6.14\cdot 10^{-3}$   & 0,0153                                & 0.0537                                \\
C12             &               &               & $2.81\cdot 10^{-4}$   & $1.47\cdot 10^{-4}$   & $1.57\cdot10^{-4}$    \\
O16             &               &               & $3.87\cdot 10^{-3}$   & $8.49\cdot 10^{-4}$   & $6.23\cdot10^{-4}$    \\
Ne20    &               &               & $1.93\cdot 10^{-5}$   & $7.03\cdot 10^{-6}$   & $1.19\cdot 10^{-5}$   \\
Mg24    &               &               & $4.91\cdot 10^{-4}$   & $8.03\cdot 10^{-4}$   & $1.17\cdot 10^{-3}$   \\
Si28    &               & 1.0   & 0.417                                 & 0.388                                 & 0.277                                 \\
S32             &               &               & 0.311                                 & 0.281                                 & 0.212                                 \\
Ar36    &               &               & 0.114                                 & 0.105                                 & 0.0916                                \\
Ca40    &               &               & 0.103                                 & 0.0924                                & 0.0845                                \\
Ti44    &               &               & $1.11\cdot 10^{-3}$   & $1.34\cdot 10^{-3}$   & $2.16\cdot 10^{-3}$   \\
Cr48    &               &               & $1.83\cdot 10^{-3}$   & $3.04\cdot 10^{-3}$   & $7.01\cdot 10^{-3}$   \\
Fe52    &               &               & $6.41\cdot 10^{-3}$   & 0.0147                                & 0.0376                                \\
Ni56    & 1.0   &               & 0.0348487                     & 0.09741397                    & 0.2324681                     \\
\hline
\end{tabular}
\end{center}
\caption{Composition of different types of ash. The ashes of type
M1, M2, and M3 are chosen to resemble NSE for  $\rho = \unit[3.2
\cdot 10^{7}]{g\ cm^{-3}}$, $\rho = \unit[5.0 \cdot 10^{7}]{g\
  cm^{-3}}$, and $\rho = \unit[1.0 \cdot 10^{8}]{g\ cm^{-3}}$.}
\label{tab:comp}
\end{table}
The ash regions were constructed by changing the composition, adding
the difference of binding energies to the internal energy of the fuel
(50\% C, 50\% O) and enforcing pressure equilibrium between the ashes
and the enviroment. The latter condition is a consequence of the
subsonic burning velocity of the deflagration front. It guarantees the
stationarity of the obstacles.

Because the EOS in FLASH does not allow the use of internal energy 
and pressure as state variables, we had to convert the 
conditions above into
initial values for temperature and density.
These are summarized in Tables\ \ref{tab:par1}. 
-\ \ref{tab:par3}.
\begin{table}[tp]
\begin{center}
\begin{tabular}{|l|*{3}{c|}}
\hline
                                                                        & N                     & S             & M1    \\
\hline 
$T$/10$^9$ K                                            &4.823          &4.639  &4.695  \\
$\rho$/10$^7$ g cm$^{-3}$                       &1.103          &1.279  &1.225  \\
$\Delta q$/10$^{17}$ erg g$^{-1}$       &7.860          &5.977  &6.273  \\
\hline 
\end{tabular}
\end{center}
\caption{Parameters of the ashes for $\rho=\unit[3.2 \cdot 10^7]{g\ cm^{-3}}$.}\label{tab:par1}
\end{table}
\begin{table}[tp]
\begin{center}
\begin{tabular}{|l|*{3}{c|}}
\hline
                                                                        & N                     & S             & M2    \\
\hline 
$T$/10$^9$ K                                            &5.505          &5.265  &5.336  \\      
$\rho$/10$^7$ g cm$^{-3}$                       &1.919          &2.205  &2.116  \\
$\Delta q$/10$^{17}$ erg g$^{-1}$       &7.860          &5.977  &6.268  \\
\hline 
\end{tabular}
\end{center}
\caption{Parameters of the ashes for $\rho=\unit[5.0 \cdot 10^7]{g\ cm^{-3}}$.}\label{tab:par2}
\end{table}
\begin{table}[tp]
\begin{center}
\begin{tabular}{|l|*{3}{c|}}
\hline
                                                                        & N                     & S             & M3    \\
\hline 
$T$/10$^9$ K                                            &6.725          &6.366  &6.357  \\      
$\rho$/10$^7$ g cm$^{-3}$                       &4.456          &5.043  &5.027  \\
$\Delta q$/10$^{17}$ erg g$^{-1}$       &7.860          &5.977  &6.007  \\
\hline 
\end{tabular}
\end{center}
\caption{Parameters of the ashes for $\rho=\unit[1.0 \cdot 10^8]{g\ cm^{-3}}$.}\label{tab:par3}
\end{table}

\subsection{One-dimensional simulations}
\begin{figure}[t]
\centerline{
\includegraphics[width = \linewidth]
  {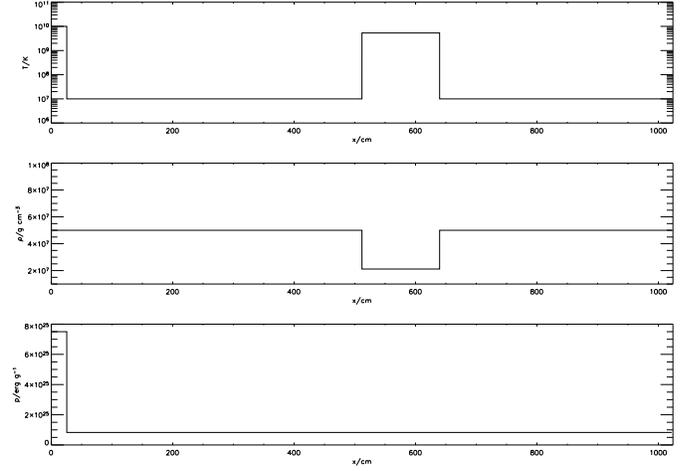}
}
\caption{Initial conditions for a system with $\rho=\unit[5.0 \cdot
  10^7]{g\ cm^{-3}}$ and an ash obstacle type M2 with width $w =
  \unit[128]{cm}$ . From top to bottom: temperature, density, pressure.}
\label{fig:setup}
\end{figure}

The initial and boundary conditions for our analyzed system are the same
as described in Sect. \ref{sec:test}. Aditionally, we add an ash obstacle
with width $w$ at the position $x = \unit[512]{cm}$
(see Fig.\ \ref{fig:setup}).
We examined the behavior of the detonation front in this setup
for widths $w = \unit[4 \dots 256]{cm}$,
compositions N,S, and M for the ash obstacle, and densities 
$\rho =\unit[3.2 \cdot 10^{7} \dots 1.0 \cdot 10^{8}]{g\ cm^{-3}}$.  

\begin{figure}[t]
\centerline{
\includegraphics[width = \linewidth]
  {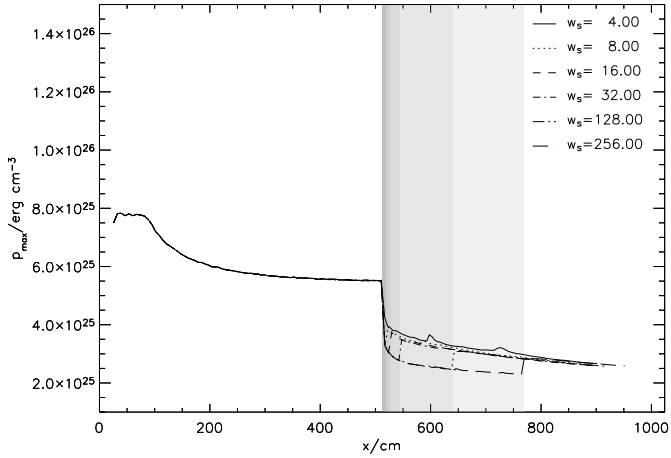}
}
\caption{Peak value of the pressure associated with the shock front
over position for $\rho=\unit[5.0 \cdot 10^7]{g\ cm^{-3}}$. The grey
shaded areas illustrate the width of the obstacle from $w=\unit[4-256]{cm}$.
The legend shows the scaled width $w_s$ of the obstacles.}
\label{fig:pmax75c}
\end{figure}

As measurement of the detonation strength, we analyzed the
peak value of the pressure associated with the shock. 
After the phase of ignition with high overpressure,
the peak value of the pressure converges to a stable state,
as can be seen in Fig.\ \ref{fig:pmax75c}. 
A stable shock front develops ahead of the reaction zone.
The peak pressure declines steeply, as soon as the shock front 
hits the obstacle at the position $x=\unit[512]{cm}$. For 
fuel densities of $\rho=\unit[3.2 \cdot 10^7]{g\ cm^{-3}}$,
$\rho=\unit[5.0 \cdot 10^7]{g\ cm^{-3}}$, and $\rho=\unit[1.0 \cdot
10^8]{g\ cm^{-3}}$ we found a reduction of peak pressure by $\unit[1.5
\cdot 10^{25}]{erg\ cm^{-3}}$, $\unit[2.5 \cdot 10^{25}]{erg\
  cm^{-3}}$, and $\unit[5.5 \cdot 10^{25}]{erg\ cm^{-3}}$.

\begin{figure}[t]
	\centering
	\begin{minipage}{0.7\linewidth}
	\subfigure[{$\rho=\unit[3{,}2 \cdot 10^7]{g\ cm^{-3}}$}, ash type N]
	{\includegraphics[width=\linewidth]{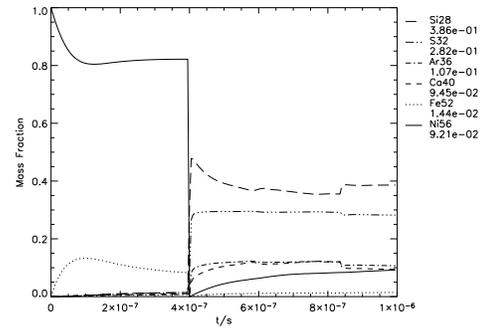}}
	\subfigure[{$\rho=\unit[5{,}0 \cdot 10^7]{g\ cm^{-3}}$}, ash type S]
	{\includegraphics[width=\linewidth]{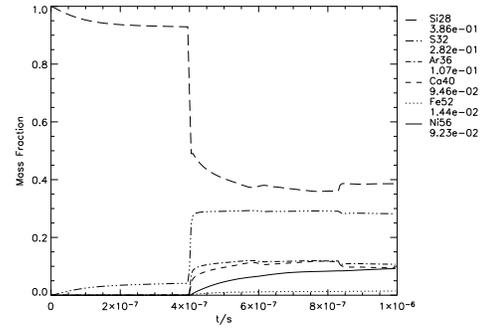}}
	\subfigure[{$\rho=\unit[1{,}0 \cdot 10^8]{g\ cm^{-3}}$}, ash type M2]
	{\includegraphics[width=\linewidth]{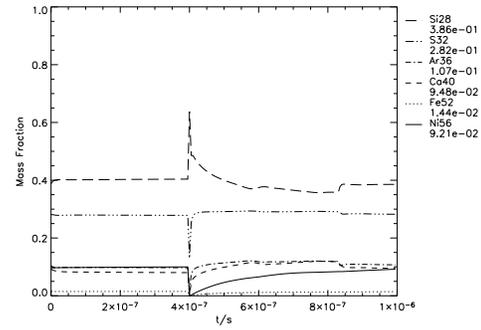}}
	\end{minipage}
	\caption{Temporal evolution of the mass fractions of the six most important isotopes
	within the obstacle for an eviromental density of $\rho=\unit[5.0 \cdot 10^7]{g\ cm^{-3}}$. 
	The legend shows the mass fractions of the specific isotopes at time
	$t=\unit[1{,}0 \cdot 10^{-6}]{s}$. }
	\label{fig:fracM}
\end{figure}

The composition seems to have virtually no influence
on the outcome. This is surprising 
because, as one can see in Fig.\ \ref{fig:fracM}, the ash is not
a inert medium. Owing to strong local variations in pressure and
temperature, its composition changes significantly during the
passage of the shock front. Nevertheless, independent of the initial composition,
the final composition is nearly identical for all types
of ash.  

\begin{figure}[t]
\centerline{
\includegraphics[width = \linewidth]
  {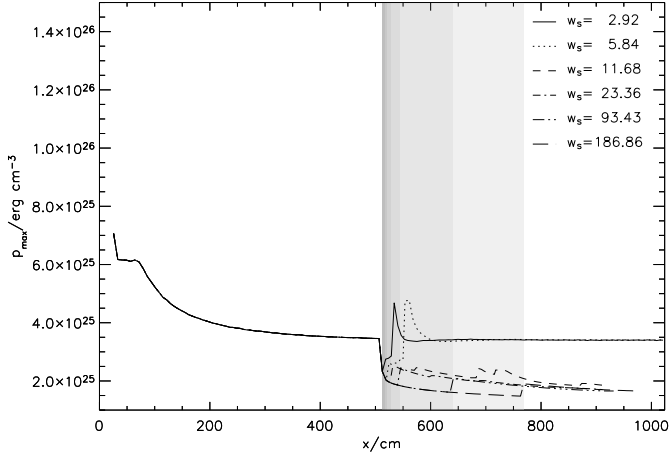}
}
\caption{Peak value of the pressure associated with the shock front
over position for $\rho=\unit[3.2 \cdot 10^7]{g\ cm^{-3}}$. The grey-shaded
areas illustrate the width of the obstacle from $w=\unit[4-256]{cm}$.
The legend shows the scaled width $w_s$ of the obstacles.}
\label{fig:pmax732c}
\end{figure}

Therefore ashes of type S and M generate nearly identical
results in the analysis of the peak pressure, but only the ash of type N leads to a 
marginally greater decrease
of the peak pressure. However, for obstacles 
with the smallest examined width $w=\unit[4]{cm}$, the decline
in the pressure maximum is approximately 30\% smaller
than in the case of a width $w=\unit[16]{cm}$ or higher. 
But this difference only seems to have consequences in the case of
the lowest analysed density of $\rho=\unit[3.2 \cdot 10^7]{g\ cm^{-3}}$,
where the detonation seems to reignite behind obstacles with
widths $w=\unit[4]{cm}$ and $w=\unit[8]{cm}$. This is shown
in Fig.\ \ref{fig:pmax732c},
where one can see that, immediately behind the 
obstacle, the pressure maximum increases
to approximately $\unit[1.0 \cdot 10^{25}]{erg\ cm^{-3}}$ above
the value ahead of the obstacle. Afterward, the peak pressure quickly converges
back to the characteristic pressure of a stable detonation. In all the other
analyzed cases, 
the peak pressure also rises abruptly behind the obstacle, but a stable
detonation front is not generated. In fact, 
the value of the peak pressure steadily decreases,
which means that the shock front continuously loses strength. The shock front
behind the obstacle is too weak to ignite a new self-sustaining detonation. 

\subsection{Two-dimensional simulations}
For the 2D simulations, we set up a system with the same length,
$l_x=\unit[1024]{cm}$, as for the 1D simulations. We set the width
of the system to $l_y=\unit[128]{cm}$ and initalized the
system with 8 top-level blocks in the $x$-direction. Because
each top-level block consists of $8 \times 8$ cells (or grid points),
this ensures that the resolution in the $x$- and $y$-directions were equal. We
used the same boundary conditions in the $x$-direction as for the 1D setup and set
periodic boundary conditions in the $y$-direction. 
As in the 1D simulation, we created an ignition area 
between $x=\unit[0]{cm}$ and $x=\unit[25.6]{cm}$ that
extended across the entire $y$-direction. The inital conditions
for the ignition area and the enviroment (composition, temperature, and velocity
of the fluid) were the same as in the 1D simulation.
 
\begin{figure}[t]
        \centering
        \subfigure[temperature $T/\lg\left(\text{K}\right)$]
        {\includegraphics[trim=0 0 33 0, angle=90,width=\linewidth,clip]
        {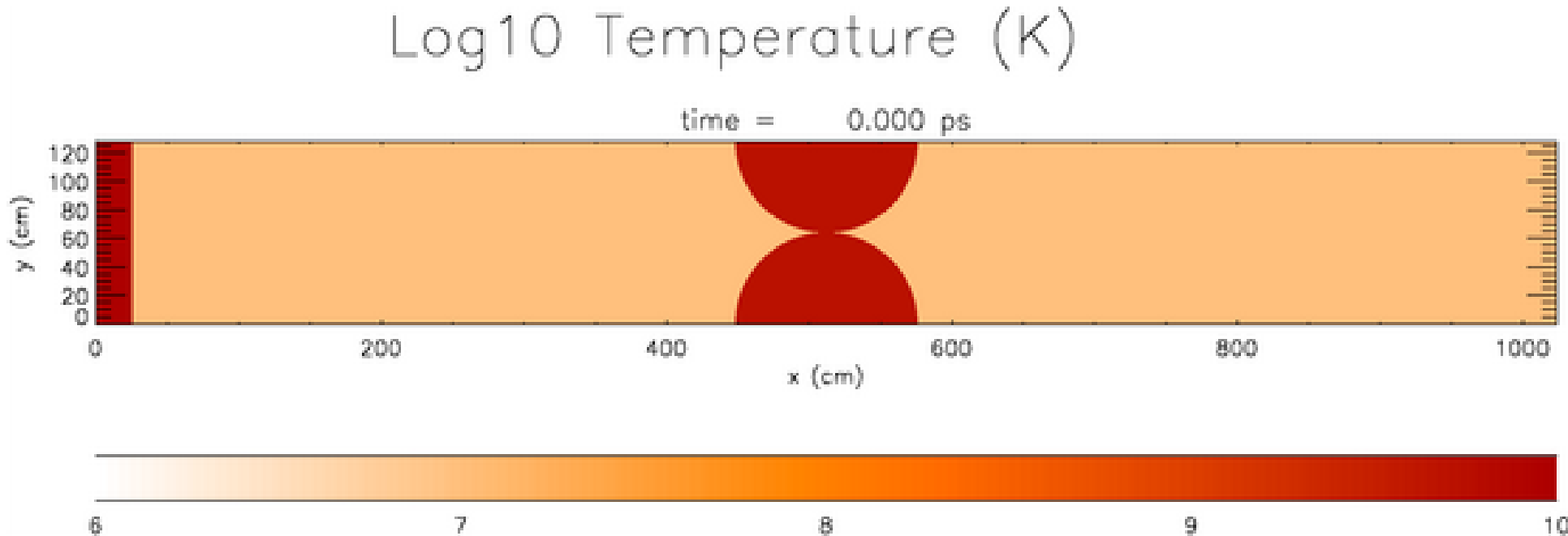}}
        \subfigure[density $\rho/\text{g cm}^{-3}$]
        {\includegraphics[trim=0 0 33 0, angle=90,width=\linewidth,clip]
        {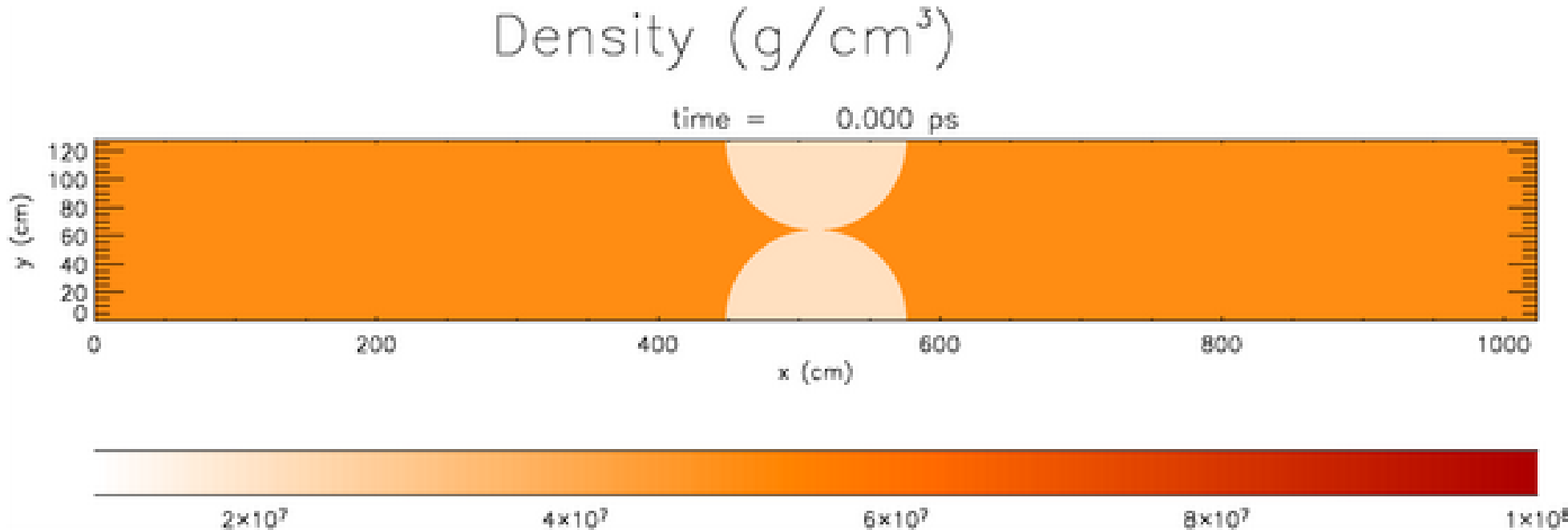}}
        \subfigure[pressure $p/\lg\left(\text{erg cm}^{-3}\right)$]
        {\includegraphics[trim=0 0 33 0, angle=90,width=\linewidth,clip]
        {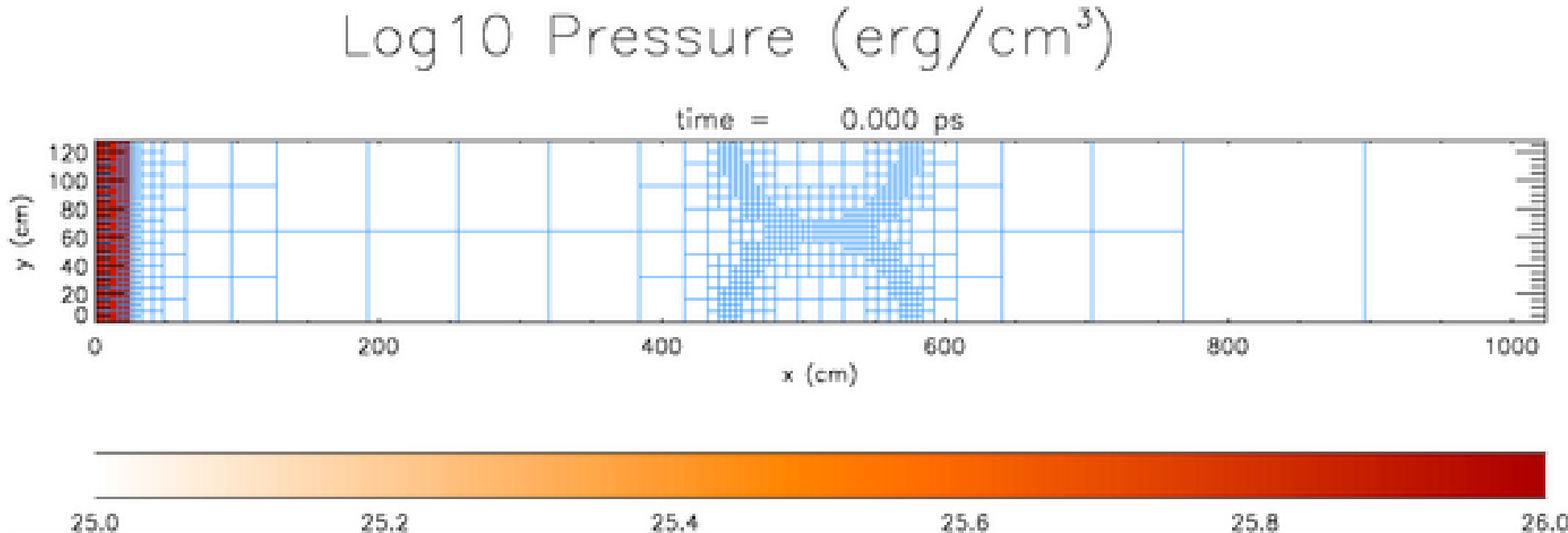}}
        \caption{Initial conditions for the 2D simulations with
        $\rho=\unit[5.0\cdot 10^7]{g\ cm^{-3}}$ and a cirular obstacle of ash 
        with radius $r=\unit[64]{cm}$. 
        }
        \label{fig:setup2d}
\end{figure}

As obstacles, we defined two semicircular areas of ash
with radius $r$, centered at the positions
$(x,y)=\unit[(512,0)]{cm}$ and $(x,y)=\unit[(512,128)]{cm}$ (equivalent to a
full circle of ash owing to the periodic boundary conditions, Fig.\
\ref{fig:setup2d}). 

We followed the evolution of the system with 
$\rho=\unit[5.0\cdot 10^7]{g\ cm^{-3}}$ and an ash obstacle of type M2
for different radii $r$. 
Limited computational resources restricted these simulations to an 
effective resolution of $\unit[0.4]{cm}$. The results of the 2D simulations,
therefore, have to be considered as preliminary.

\begin{figure}[tp]
        \centering
        \includegraphics
        [bb=330 100 518 662, trim=50 0 33 0, angle=90,width=\linewidth,clip]
        {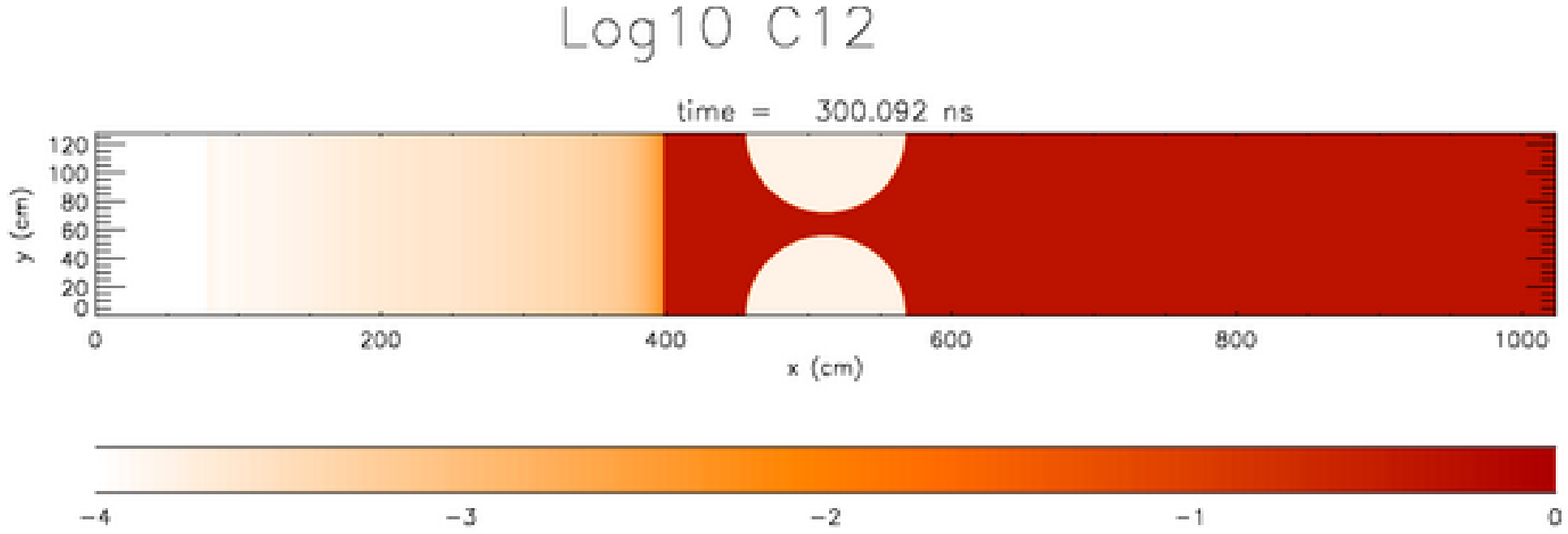}
        \includegraphics
        [bb=330 100 518 662, trim=50 0 33 0, angle=90,width=\linewidth,clip]
        {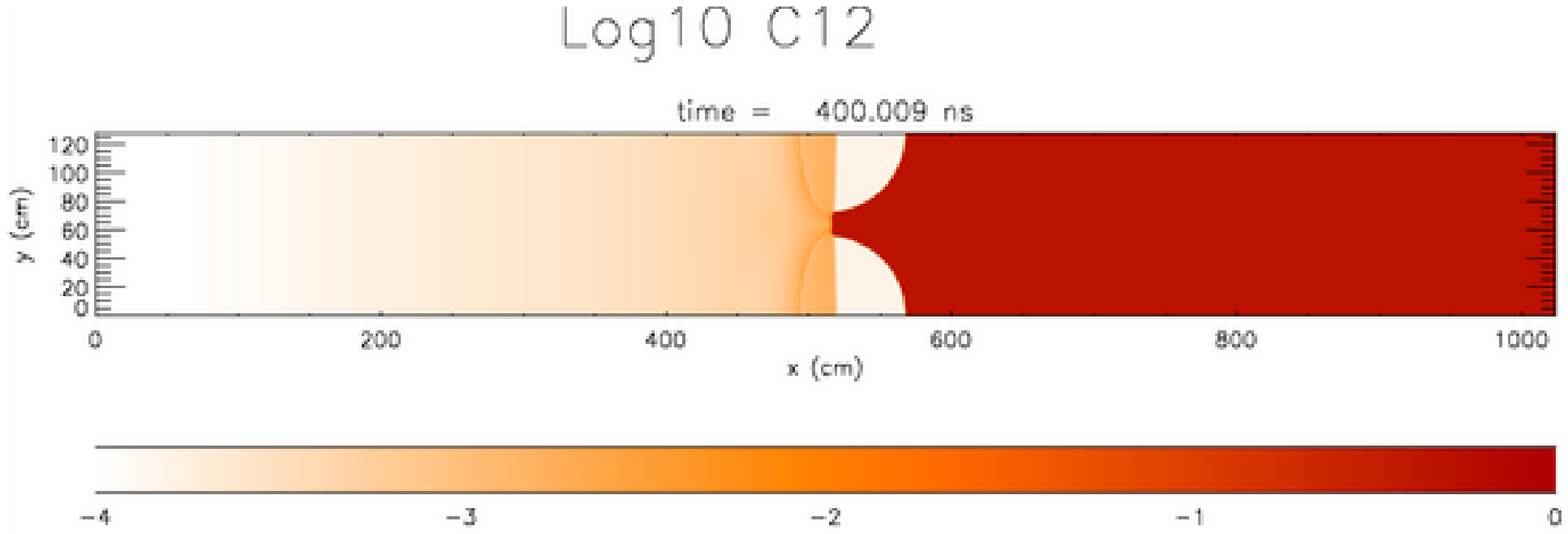}
        \includegraphics
        [bb=330 100 518 662, trim=50 0 33 0, angle=90,width=\linewidth,clip]
        {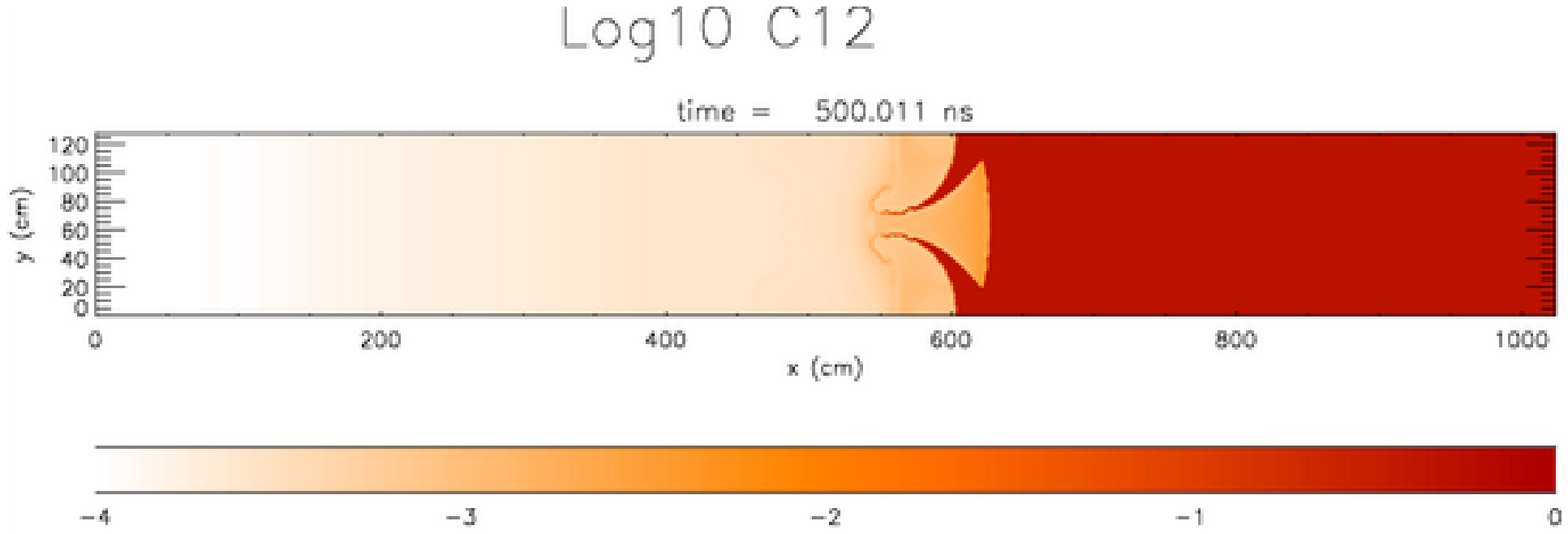}
        \includegraphics
        [bb=330 100 518 662, trim=50 0 33 0, angle=90,width=\linewidth,clip]
        {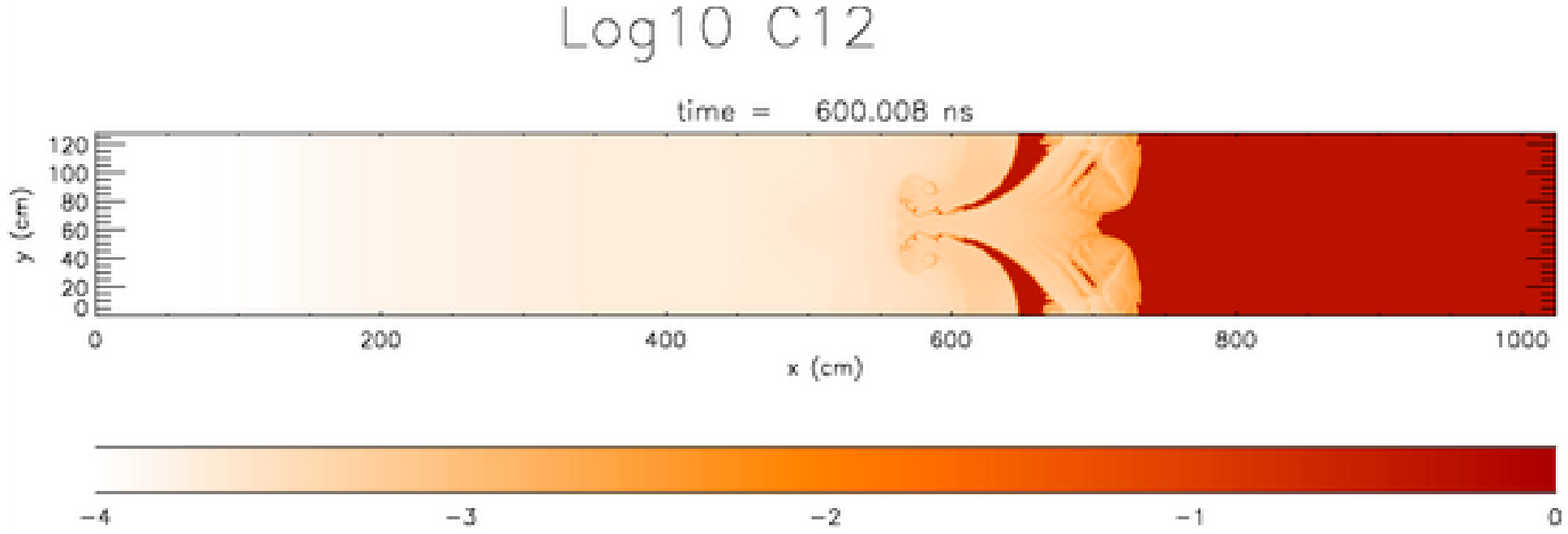}
        \includegraphics
        [bb=330 100 518 662, trim=50 0 33 0, angle=90,width=\linewidth,clip]
        {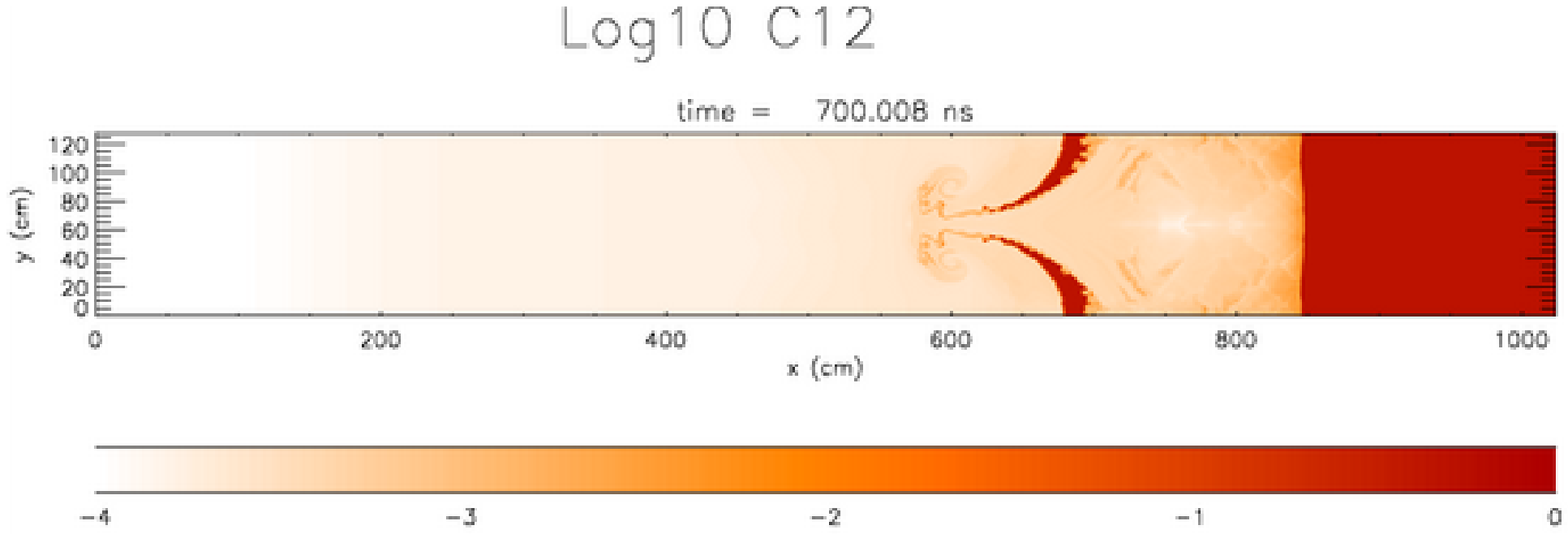}
        \includegraphics
        [bb=330 100 518 662, trim=0 0 33 0, angle=90,width=\linewidth,clip]
        {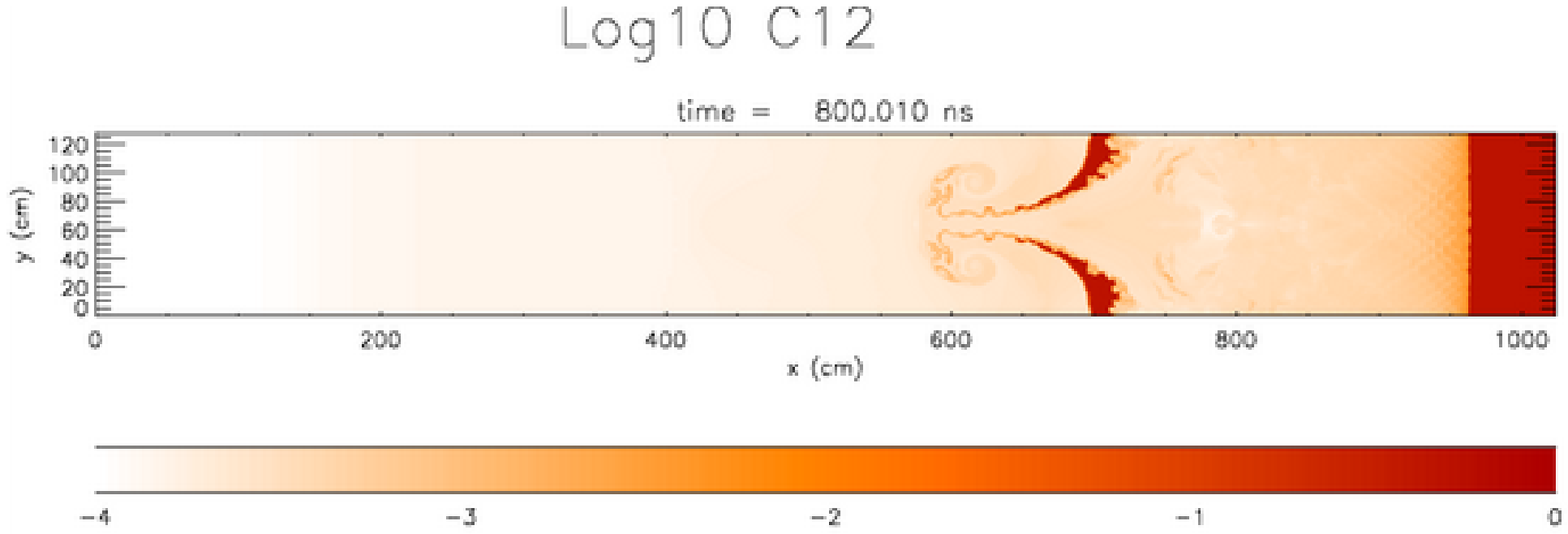}
        \caption{Carbon mass fraction using logarithmic scaling. 
        The distance between the two semicircles is $\unit[16]{cm}$.}
        \label{fig:c1216}
\end{figure}

Figure\ \ref{fig:c1216} shows the development of the carbon mass fraction for a detonation
front hitting a circular obstacle of ash with radius $r=\unit[56]{cm}$,
i.e., the distance $a$ between the two semicircles of ash is
$\unit[16]{cm}$. 
\begin{figure}[tp]
        \centering
        \includegraphics
        [bb=330 100 518 662, trim=50 0 33 0, angle=90,width=\linewidth,clip]
        {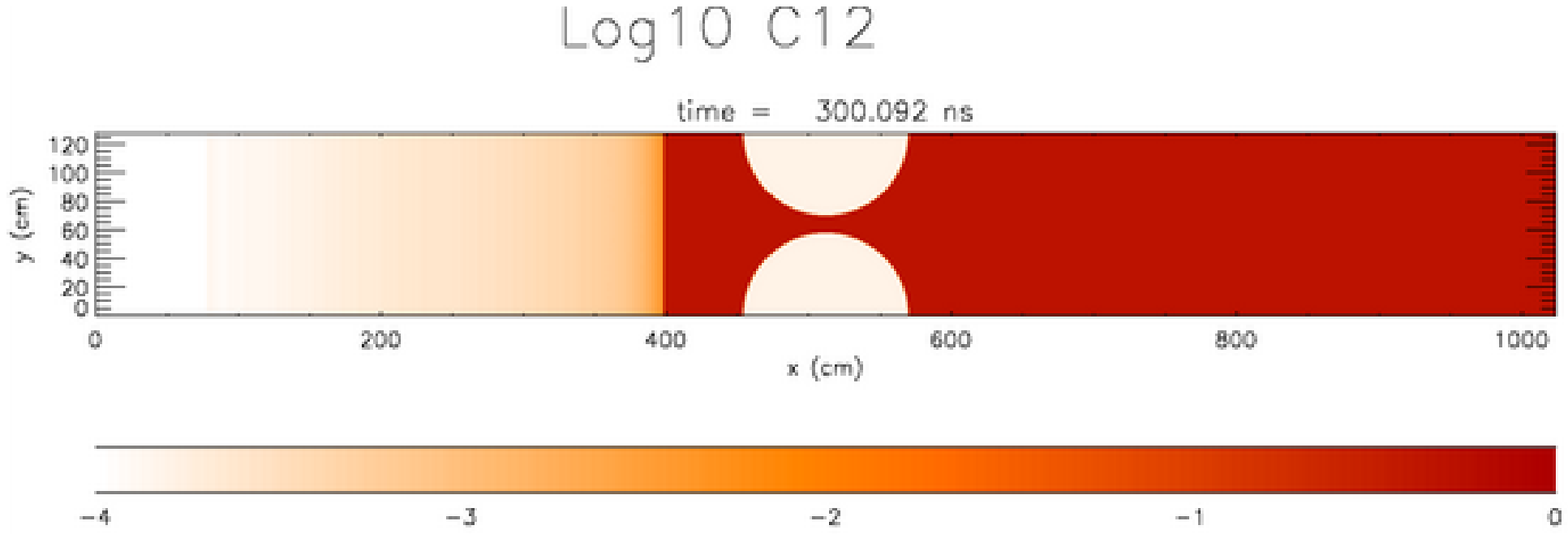}
        \includegraphics
        [bb=330 100 518 662, trim=50 0 33 0, angle=90,width=\linewidth,clip]
        {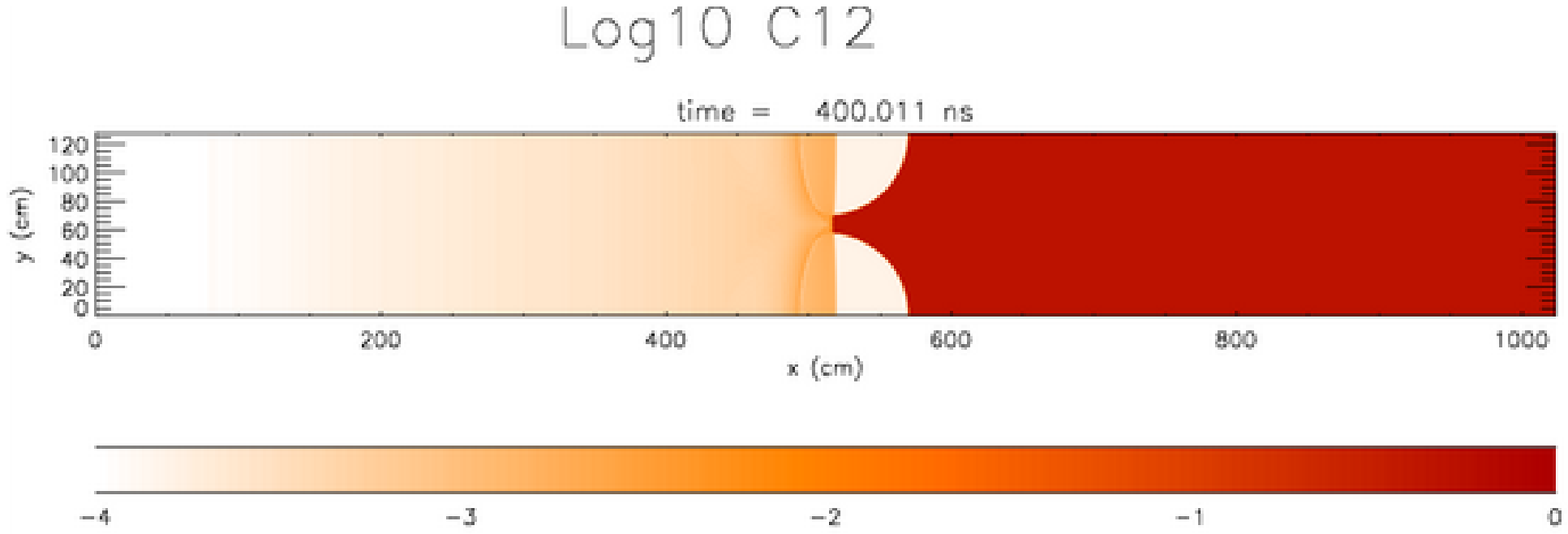}
        \includegraphics
        [bb=330 100 518 662, trim=50 0 33 0, angle=90,width=\linewidth,clip]
        {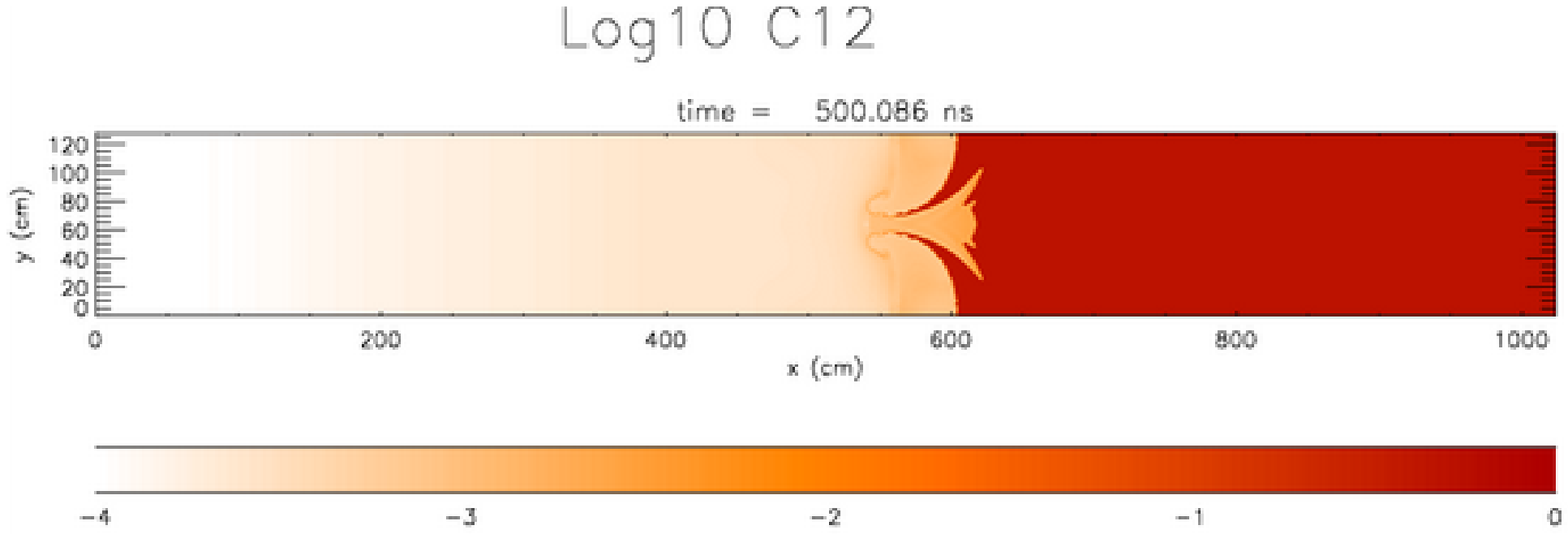}
        \includegraphics
        [bb=330 100 518 662, trim=50 0 33 0, angle=90,width=\linewidth,clip]
        {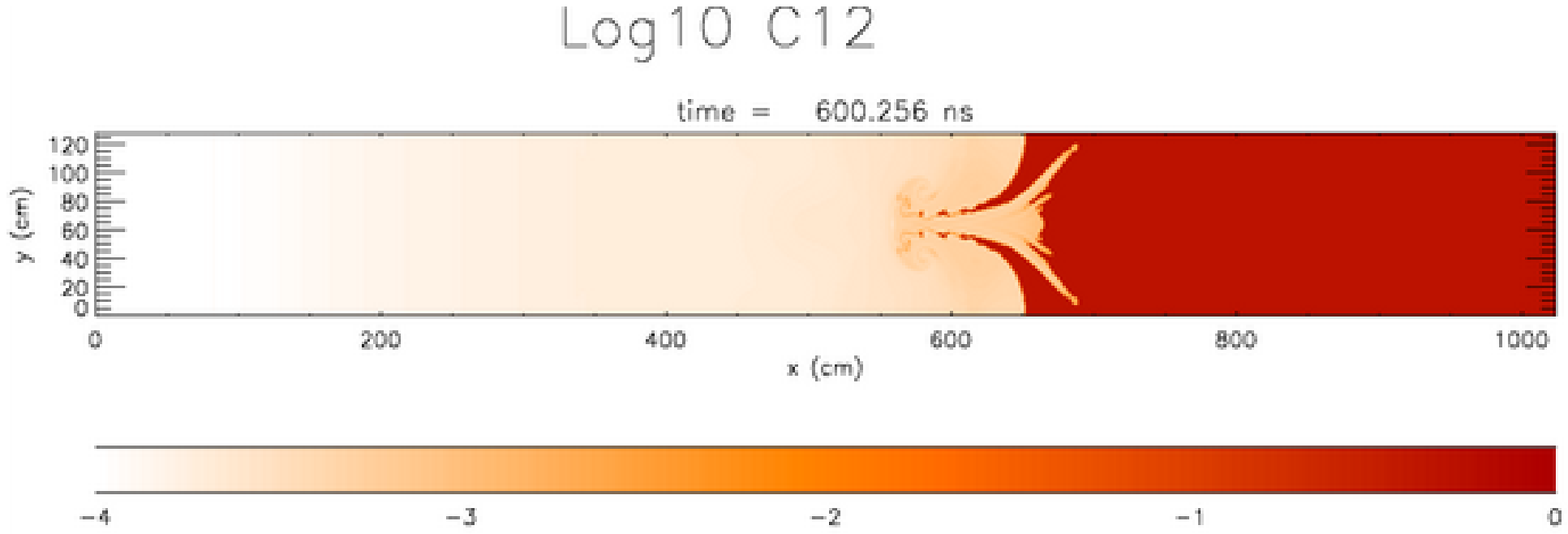}
        \includegraphics
        [bb=330 100 518 662, trim=50 0 33 0, angle=90,width=\linewidth,clip]
        {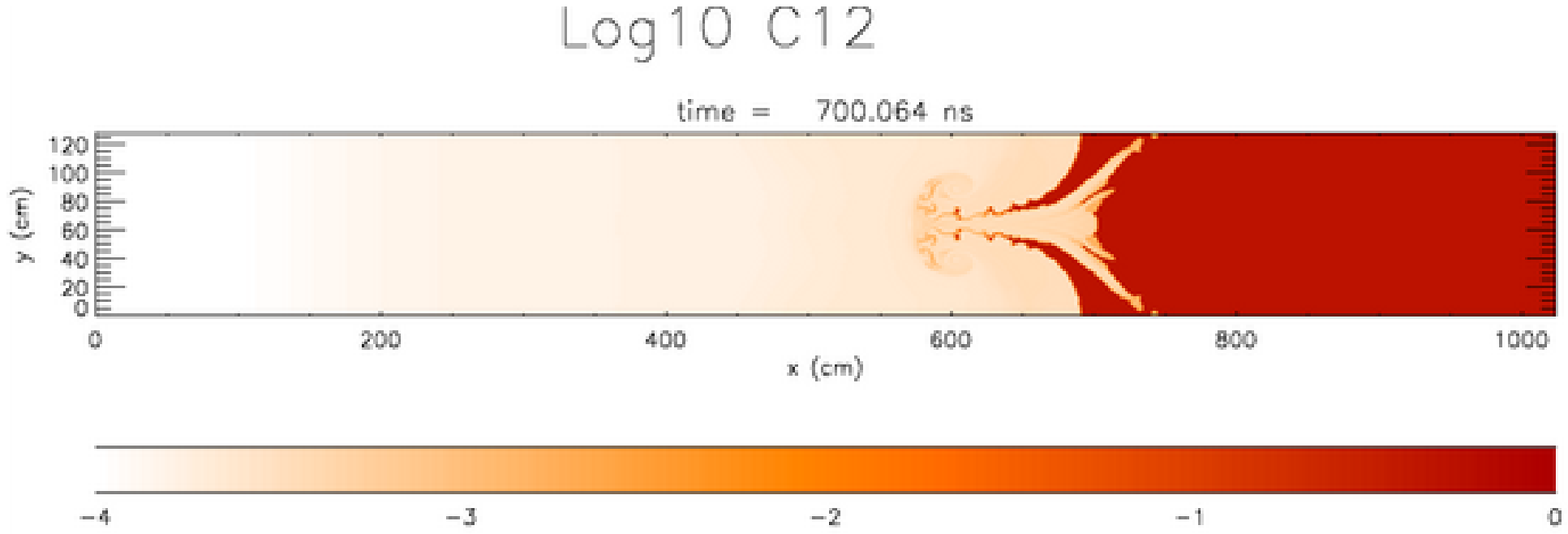}
        \includegraphics
        [bb=330 100 518 662, trim=0 0 33 0, angle=90,width=\linewidth,clip]
        {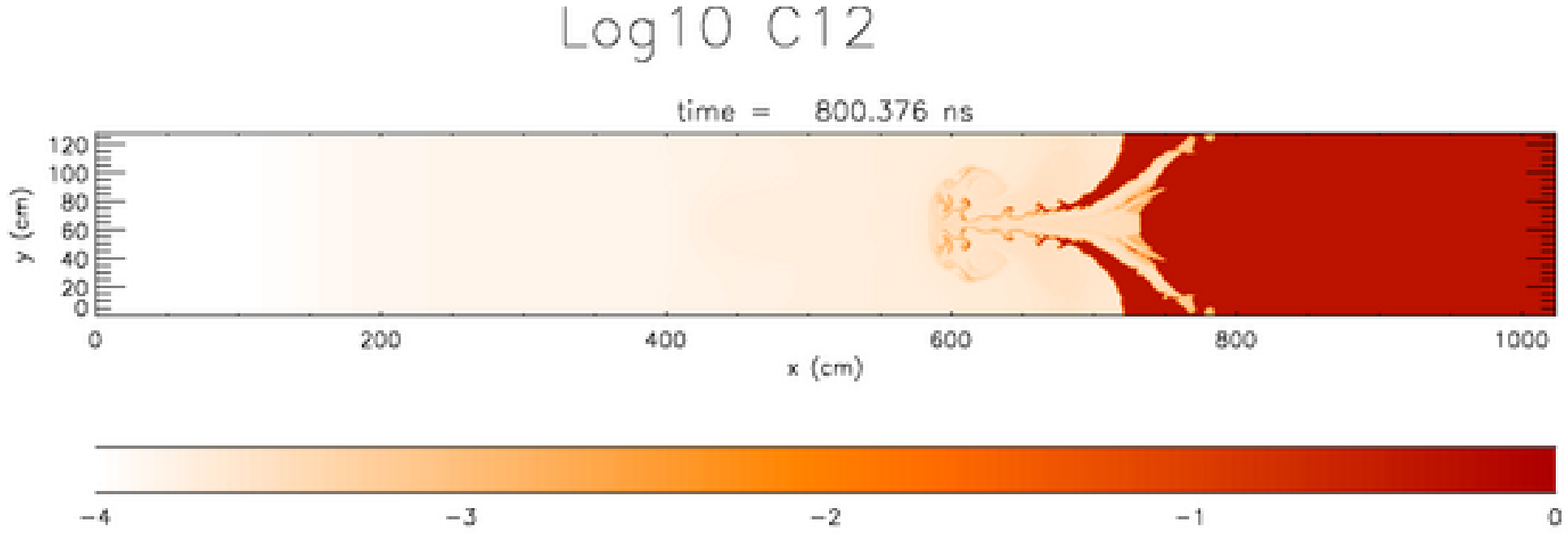}
        \caption{Carbon mass fraction using logarithmic scaling. 
        The distance between the two semicircles is $\unit[12]{cm}$.}
        \label{fig:c1212}
\end{figure}  
One can see that the detonation squeezes between 
the two semicircles and re-expands after passing the narrowest point of the
system. However, a ``shadow'' of unburned material remains behind the former
obstacle of ash. Figure\ \ref{fig:c1216} also shows
a mushroom-like structure produced by the Richtmyer-Meshkov
instability \citep{Ri60,Mesh70} occuring when a shock
front passes the contact discontinuity between two fluids of different
densities. Nethertheless, the detonation front is not stopped by an
obstacle of these proportions. 

A different result was found for the distance $a \leq \unit[12]{cm}$, as can
be seen in Fig.\ \ref{fig:c1212}.
Although the detonation front propagates through the small funnel 
between the two semicircles of ash, it becomes unstable and dies
right behind the obstacle. Apparently, the funnel of unburned material
between the two obstacles is too small to provide enough
energy for a self-sustaining detonation front. 

\section{Conclusions}

Our results have immediate consequences for models of thermonuclear
detonations in large-scale SN Ia simulations. In addition to an accurate
reproduction of the detonation speed, which is usually achieved by
simulating an unresolved detonation that is directly coupled to
the nuclear reactions via the coarse-grained temperature, we must
enforce the new requirement that the front stops immediately when it
encounters burned material, since the typical size of ash regions that
it may cross can never be resolved in practice. Whether or not unresolved,
temperature-coupled detonations (frequently used in
multi-dimensional explosion models on large scales) are able to meet
this requirement will
have to be demonstrated for each numerical implemetation. The
level-set algorithm for detonations has been shown to
handle this task accurately\citep{GN05}.

Another lesson learned from our simulations is the necessity to fully
resolve the thermal detonation profile in order to obtain an accurate
description of detonation quenching and ignition. If the spatial
resolution is chosen too coarsely in Eulerian methods, numerical
diffusion of heat from the 
ash pre-conditions the fuel and helps to re-ignite the detonation
(1D Lagrangian codes are not subject to this restriction). We
have found that mildly underresolved detonations were able to survive
the passage of large regions of ash that turned to out to be a
numerical artifact under closer scrutiny. 

Theoretical predictions for SNe Ia spectra depend on ejecta
inhomogeneities over a wide range of length scales, whereas this work
only focussed on the microscopic regime. A comparison of
3D simulations of pure deflagration and 
delayed detonation models for SNe Ia will reveal the ability of
delayed detonations to enhance the chemical stratification of the
ejecta and to reduce the amount of unburnt material at low
velocities. This investigation is currently in progress.

\acknowledgements
We would like to thank Marcus Brüggen and Pawel Ciecielag for their
help with the FLASH code, and Christian Klingenberg, Irina Golombek,
Ewald Müller, Jan Pfannes, and Wolfram Schmidt for helpful discussions. 
The research of
JCN was supported by the Alfried Krupp Prize for Young 
University Teachers of the Alfried Krupp von Bohlen und Halbach Foundation.
The software used in this work was in part developed by the DOE-supported 
ASC / Alliance Center for Astrophysical Thermonuclear Flashes at the 
University of Chicago.
\bibliographystyle{aa}
\bibliography{snrefs,supernova2}

\end{document}